\documentclass[twocolumn,showpacs,amsmath,amssymb,pra,superscriptaddress]{revtex4}
\usepackage{graphicx}
\usepackage{bm}
\usepackage{dcolumn}
\begin{document}
%\title{Comment on "Quantum mechanics of a constrained electrically charged particle in the presence of electrical currents''}
\title{Absence of anomalous interactions in the quantum theory of constrained charged particles in presence of electrical currents}
\author{Carmine  Ortix}
\affiliation{Institute Lorentz for Theoretical Physics, Leiden University, Leiden, The Netherlands}
\affiliation{Leibniz-Institute for Solid-State and Materials Research, IFW Dresden, Germany}
 \author{Jeroen van den Brink}
 \affiliation{Institute Lorentz for Theoretical Physics, Leiden University, Leiden, The Netherlands}
\affiliation{Leibniz-Institute for Solid-State and Materials Research, IFW Dresden, Germany}
\date{\today}

\begin{abstract}
The experimental progress in synthesizing low dimensional nanostructures where carriers are confined to bent surfaces has boosted the interest in the theory of quantum mechanics on curved two-dimensional manifolds. The current theoretical paradigm relies on a thin wall quantization method introduced by Da Costa \cite{dac82}. In Ref.~[\onlinecite{jen09}] Jensen and Dandoloff  find that when employing the thin wall quantization in presence of externally applied electric and magnetic fields, an unphysical orbital magnetic moment appears. This anomalous result is taken as evidence for the breakdown of the thin wall quantization. We show, however, that in a proper treatment within the Da Costa framework: ($i$) this anomalous contribution is absent and ($ii$) there is no coupling between external electromagnetic field and surface curvature \cite{footnote1}. 
\end{abstract}

\maketitle

To derive the thin wall quantization in presence of external electromagnetic field we start with the Schr\"odinger equation that is minimally coupled with the four component vector potential in a generic curved three-dimensional space.  Adopting Einstein summation convention and tensor covariant and contravariant components, we have
\begin{eqnarray}
i \hbar \left[\partial_t-\dfrac{i Q A_0}{\hbar}\right]  \psi&=& -\dfrac{\hbar^2} {2 m} G^{i j}  \left[{\cal D}_i-\dfrac{i Q  A_i}{\hbar}\right] \times \nonumber \\ & &  \left[{\cal D}_j-\dfrac{i Q A_j}{\hbar}  \right] \psi 
\label{eq:scheq1}
\end{eqnarray}
where $Q$ is the particle charge, $G^{i j}$ are the contravariant components of the metric tensor and  $A_i$ are the covariant components of the vector potential ${\bf A}$ with the scalar potential defined by $V=-A_0$.
The covariant derivative ${\cal D}_i$ is as usual defined by
$${\cal D}_i v_j=\partial_i v_j - \Gamma_{i j}^k v_k,$$ 
where $v_j$ are the covariant components of a vector field ${\bf v}$ and
 $\Gamma_{i j}^k$ are the Christoffel symbols
$$\Gamma_{i j}^k= \dfrac{1}{2} G^{k l} \left[\partial_j G_{l i}+\partial_i G_{l j} -\partial_l G_{i j}\right].$$ 
From Eq.~(\ref{eq:scheq1}) one obtains
\begin{eqnarray}
i \hbar {\cal D}_0  \psi&=&-\dfrac{1}{2 m} \left[ \hbar^2 G^{i j} \, \partial_i \,\partial_j \,\psi-\hbar^2\, G^{i j} \,\Gamma_{i j}^k \,\partial_k \,\psi \right. \nonumber \\ & & \left.- i \,Q \,\hbar\, G^{ij}\, \partial_i \,\left(A_j \psi\right)+ i\,Q \,\hbar\, G^{ij}\, \Gamma_{ij}^k \,A_k \,\psi  \right.\nonumber \\ & & \left. -i\, Q \,\hbar\, G^{ij}\, A_i \,\partial_j\, \psi -Q^2\, G^{ij}\, A_i\, A_j \,\psi \right], \label{eq:scrogenericexpanded}
\end{eqnarray}
where for convenience we introduced the time gauge covariant derivative ${\cal D}_0=\partial_t - i Q A_0 / \hbar$. To proceed further, it is useful to define a coordinate system. As in Ref.~\cite{dac82,fer08,jen09} we consider a surface ${\cal S}$ with parametric equations ${\bf r}={\bf r}(q_1,q_2)$. The portion of the 3D space in the immediate neighborhood of ${\cal S}$ can be then parametrized as ${\bf R}(q_1,q_2)={\bf r}(q_1,q_2)+q_3 {\hat N}(q_1,q_2)$ with ${\hat N}(q_1,q_2)$ the unit vector normal to ${\cal S}$. We then find, in agreement with previous studies \cite{dac82,fer08,jen09}, the relations among $G_{ij}$ and the covariant components of the 2D surface metric tensor $g_{i j}$ to be
$$G_{ij}=g_{ij}+\left[\alpha \, g + \left(  \, \alpha g \right)^{T}\right]_{ij} q_3+ \left( \alpha \, g \alpha^T \right)_{ij} \, q_3^2 \hspace{.5cm} i,j=1,2$$
$$G_{i,3}=G_{3,i}=0 \hspace{.5cm} i=1,2; \hspace{1cm} G_{3,3}=1$$
with $\alpha$ indicating the Weingarten curvature tensor of the surface ${\cal S}$ \cite{dac82,fer08}. We recall that 
the mean curvature $M$ and the Gaussian curvature $K$ of the surface ${\cal S}$ are related to the Weingarten curvature tensor by
\begin{eqnarray*}
M&=& \dfrac{{\it Tr} (\alpha)}{2} \\
K&=& {\it Det} (\alpha)
\end{eqnarray*} 

Now we can apply the thin-layer procedure introduced by Da Costa \cite{dac82} and take into account the effect of a confining potential $V_{\lambda}(q_3)$, where $\lambda$ is a squeezing parameter that controls the strength of the confining potential. When $\lambda$ is large, the total wavefunction will be localized in a narrow range close to $q_3=0$. This allows one to take the $q_3 \rightarrow 0$ limit in the Schr\"odinger equation. For the sake of clarity, it is better to introduce the indices $a,b$ which can assume the values $1,2$ alone. From the structure of the metric tensor, one finds the following limiting relations:
\begin{eqnarray*}
\Gamma_{3 3}^a & \equiv & 0 \\
\lim_{q_3 \rightarrow 0} G^{a b} \Gamma_{a b}^3 & \equiv & - 2 M
\end{eqnarray*}
With this, the Schr\"odinger equation Eq.~(\ref{eq:scrogenericexpanded}) can be then recast as
$$i \hbar {\cal D}_0 \psi={\cal H}_{\cal S} \psi + {\cal H}_{N} \psi$$
where ${\cal H}_{\cal S}$ is an effective 2D Hamiltonian that depends on the surface {\cal S} alone and reads
\begin{eqnarray*}
{\cal H}_{\cal S}&=&-\dfrac{1}{2 m} \left[ \hbar^2 g^{a b} \, \partial_a \,\partial_b \,\psi-\hbar^2\, g^{a b} \,\Gamma_{a b}^k \,\partial_k \,\psi \right. \nonumber \\ & & \left.- i \,Q \,\hbar\, g^{a b}\, \partial_a \,\left(A_b \psi\right)+ i\,Q \,\hbar\, g^{a b}\, \Gamma_{a b}^k \,A_k \,\psi  \right.\nonumber \\ & & \left. -i\, Q \,\hbar\, g^{a b}\, A_a \,\partial_b\, \psi -Q^2\, g^{a b}\, A_a\, A_b \,\psi \right].
\end{eqnarray*}
The Hamiltonian ${\cal H}_{N}$ that depends on the degrees of freedom in the  normal direction to the surface $S$ is 
\begin{eqnarray}
{\cal H}_{N}&=&-\dfrac{1}{2 m} \left[\hbar^2 \partial_3^2 \psi+ 2  \hbar^2 M \partial_3 \psi- i \, Q \, \hbar \partial_3 (A_3 \psi) \right. \nonumber \\
& & \left. - 2 i \, Q \, \hbar  M  A_3 \psi -i \, Q \, \hbar   A_3 \partial_3 \psi- \, Q^2 A_3^2 \psi  \right], \label{eq:hamiltonianq3}
\end{eqnarray}
where we left out the confining potential term. In the equation above, the linear coupling between the $A_3$ component of the vector potential and the mean curvature of the surface through the term $Q \, M  A_3$ corresponds to the anomalous contribution to the orbital angular moment of the charged particle found in Ref.~\cite{jen09}. Now we show that this term identically cancels by considering the effective Schr\"odinger equation for a well-defined surface wavefunction. In agreement with  Ref.~\cite{fer08}, we subsequently find that there is no coupling between an external magnetic field and the curvature of the surface, independent of the shape of the surface. 

In order to obtain a surface wavefunction depending on $(q_1,q_2)$ alone \cite{dac82,fer08,jen09}, we introduce a new wavefunction $\chi(q_1,q_2,q_3)$ that in the case of separability decomposes as $\chi(q_1,q_2,q_3)=\chi_{\cal S} (q_1,q_2) \times \chi_{N} (q_3)$.  
Conservation of the norm requires
$$\psi(q_1,q_2,q_3)=\left[1+ 2 M q_3+ K  q_3^2 \right]^{-1/2} \chi(q_1,q_2,q_3).$$
In the $q_3 \rightarrow 0$ limit, the following relations hold:
\begin{eqnarray*}
\lim_{q_3 \rightarrow 0} \psi &=& \chi \\
\lim_{q_3 \rightarrow 0} \partial_3 \psi&=& \partial_3 \chi - M \chi \\
\lim_{q_3 \rightarrow 0} \partial^2_3 \psi&=& \partial_3^2 \chi- 2 M  \partial_3 \chi+ 3 M^2 \chi - K  \chi. 
\end{eqnarray*} 
With these limiting relations,  ${\cal H}_{N}$ reduces to
\begin{eqnarray}
{\cal H}_{N}&=&-\dfrac{\hbar^2}{2 m} \left[\partial_3^2 \chi+ \left(M^2 - K \right) \chi- 2 \, i \,Q \,\hbar\, A_3 \, \partial_3 \chi \right. \nonumber \\ & & \left. - i \,Q \,\hbar \,\chi \,\partial_3 A_3- Q^2\, A_3^2\, \chi\right]
\end{eqnarray}
The second term in the square brackets corresponds to Da Costa's curvature induced scalar potential \cite{dac82}. 

The Hamiltonian ${\cal H}_{N}$ governing the dynamics along the direction normal to the surface is thus coupled to the dynamics on the surface $S$ only via the term $A_3 \partial_3 \chi$ \cite{fer08}. One finds that, contrary to Ref.~[\onlinecite{jen09}],  a coupling between the surface curvature and the external vector potential is lacking. This implies absence of the fundamental problem that was signaled in Ref.~[\onlinecite{jen09}] in Da Costa's thin wall quantization method -- instead the method is well-founded and can be employed without restriction.

This research is supported by the Dutch Science Foundation FOM.


\begin{thebibliography}{1}

\bibitem{footnote1}
This Comment was communicated to the authors of Ref.~\cite{jen09}, B. Jensen and R. Dandoloff, on December 10th 2009.

\bibitem{dac82}
R.C.T. da Costa, Phys. Rev. A {\bf 23},  1982  (1981).

\bibitem{jen09}
B. Jensen and R. Dandoloff, Phys. Rev. A  {\bf 80},  052109  (2009).

\bibitem{fer08}
G. Ferrari and G. Cuoghi, Phys. Rev. Lett. {\bf 100},  230403  (2008).

\end{thebibliography}
\end{document}